\documentclass[sigconf,screen]{acmart-new}
\usepackage{amsmath,amsfonts}
\usepackage{algorithmic}
\usepackage{graphicx}
\usepackage{textcomp}
\usepackage{xcolor}
\usepackage{algorithm2e}
\usepackage{multirow}

\usepackage{booktabs}
\usepackage{caption}
\usepackage{subcaption}

\usepackage{verbatim}
\usepackage{makecell}
\usepackage{lstautogobble}
\usepackage{listings}
\usepackage{enumitem}

\renewcommand{\paragraph}[1]{\vspace{5pt}\noindent\textbf{#1.}}

\usepackage{cleveref}
\crefname{chapter}{Chapter}{}
\crefname{section}{Section}{Sections}
\crefformat{section}{Section~#2#1#3}

\crefname{table}{Table}{Tables}
\crefname{figure}{Figure}{Figures}
\crefname{algorithm}{Alg.}{Algs.}
\crefname{line}{Line}{Lines}
\crefname{appendix}{App.}{}
\crefname{chapter}{Chapter}{Chapters}

\crefname{thm}{Theorem}{Theorems}
\crefname{prop}{Proposition}{Propositions}
\crefname{definition}{Definition}{Definitions}
\crefname{lemma}{Lemma}{Lemmas}
\crefname{cor}{Corollary}{Corollaries}
\crefname{equation}{Eq.}{Eqs.}

\creflabelformat{equation}{#2\textup{#1}#3}
\crefrangelabelformat{equation}{(#3#1#4--#5#2#6)}

\def\BibTeX{{\rm B\kern-.05em{\sc i\kern-.025em b}\kern-.08em
    T\kern-.1667em\lower.7ex\hbox{E}\kern-.125emX}}

\allowdisplaybreaks

\setcopyright{acmlicensed}
\acmDOI{10.1145/3617572.3617881}
\acmYear{2023}
\copyrightyear{2023}
\acmSubmissionID{fsews23sddmain-p59-p}
\acmISBN{979-8-4007-0377-5/23/12}
\acmConference[SDD '23]{Proceedings of the 1st International Workshop on Software Defect Datasets}{December 8, 2023}{San Francisco, CA, USA}
\acmBooktitle{Proceedings of the 1st International Workshop on Software Defect Datasets (SDD '23), December 8, 2023, San Francisco, CA, USA}
\received{2023-07-31}
\received[accepted]{2023-08-21}
    
\begin{document}
\SetKwInput{KwInp}{Hyperparameters}
\SetKw{KwIn}{in}

\title{Log Summarisation for Defect Evolution Analysis}

\author{Rares Dolga}
\orcid{0000-0002-1800-411X}
\affiliation{%
  \institution{JPMorgan Chase}
  \city{}
  \country{UK}
}
\email{rares.dolga@jpmchase.com}

\author{Ran Zmigrod}
\orcid{0009-0001-0169-1670}
\affiliation{%
  \institution{JPMorgan Chase}
  \city{}
  \country{UK}
}
\email{ran.zmigrod@jpmorgan.com}

\author{Rui Silva}
\orcid{0009-0003-6016-6432}
\affiliation{%
  \institution{JPMorgan Chase}
  \city{}
  \country{UK}
}
\email{rui.silva@jpmorgan.com}

\author{Salwa Alamir}
\orcid{0009-0006-6650-7041}
\affiliation{%
  \institution{JPMorgan Chase}
  \city{}
  \country{UK}
}
\email{salwa.alamir@jpmchase.com}

\author{Sameena Shah}
\orcid{0009-0000-5960-5811}
\affiliation{%
  \institution{JPMorgan Chase}
  \city{}
  \country{UK}
}
\email{sameena.shah@jpmchase.com}

\begin{abstract}
Log analysis and monitoring are essential aspects in software maintenance and identifying defects.
In particular, the temporal nature and vast size of log data leads to an interesting and important research question: How can logs be summarised and monitored over time?
While this has been a fundamental topic of research in the software engineering community, work has typically focused on heuristic-, syntax-, or static-based methods.
In this work, we suggest an online semantic-based clustering approach to error logs that dynamically updates the log clusters to enable monitoring code error life-cycles.
We also introduce a novel metric to evaluate the performance of temporal log clusters.
We test our system and evaluation metric with an industrial dataset and find that our solution outperforms similar systems.
We hope that our work encourages further temporal exploration in defect datasets.

\end{abstract}

\ccsdesc[500]{Theory of computation~Unsupervised learning and clustering}
\ccsdesc[500]{Software and its engineering~Software defect analysis}

\keywords{Defect Detection, Log Analysis, Clustering, NLP}

\maketitle

\section{Introduction}
Log data is crucial in the analysis, maintenance, and fixing of errors in software systems \cite{bug_localisation, post-mortem}.
Logs are verbose documents that detail a system's state throughout its execution lifecycle and as such, lead to vast amounts of data that make manual analysis intractable in practice \cite{clust_survey}.
Consequently, both research and industry communities have focused on automatic log analysis tasks such as failure diagnosis and prediction \cite{failDiag}, anomaly detection \cite{anom}, log comprehension \cite{logComprehension}, \emph{inter alia}.

Log data for a system is temporal, i.e., logs stream sequentially as a system is developed and tested.
Therefore, one can monitor the existence and persistence of code defects by monitoring their existence in logs \cite{FLAP}.
To do this, past research has focused on focused log comprehension \cite{cyber_survey, log_anomaly} and log evolution tracking.
The latter task has mainly been approached with heuristic-based systems \cite{rule_based1, LenMa} and unsupervised syntax-based systems \cite{LogMine,DRAIN,towards_detecting_patterns, starlink}.
These systems work well for small domains where data is structured, however, the unstructured and large nature of modern log data limit the success of these techniques for real world data.

Another approach to capture log evolution is to consider the semantic representations of logs.
Using semantic rather than syntactic meaning for logs makes sense as they may vary across applications and developers (i.e., have different free-form text).
Semantic representations for logs have been previously used for anomaly detection and defect prediction due to the free-text nature of logs \cite{swiss_log, log_anomaly}.
Moreover, past work has also focused on static semantic log clustering using common approaches (e.g.,  K-Means or Gaussian Mixture Models) \cite{clust_error_mess}.
While dynamic clustering techniques exist \cite{denStream}, they lead to overly erratic log evolution.
Therefore, a gap exists in the software defect tracking literature that we aim to explore.

In this paper, we introduce a novel online algorithm for clustering and monitoring logs based on their semantic representation.
Through collaboration with a team of $61$ Software Reliability Engineers (SREs), who are responsible for almost $500$ applications that jointly produce roughly $100,\!000$ logs per day, we extracted three key criteria for successful monitoring of log evolution: identifying errors, providing meaningful representative messages, and minimising disruption of log clusters.
These criteria were used to construct a novel evaluation metric that captures real world performance for temporal log clustering.
Experiments using a private industrial dataset demonstrate that our online clustering method outperforms existing static and dynamic clustering methods.
Additionally, we demonstrate that using a richer semantic representation further improves performance.
We note that our work could be used in generating more high level error log topics in both temporal and non-temporal datasets.
The contribution of this works are:
\begin{enumerate}[leftmargin=0.6cm]
    \item We introduce a novel online semantic-based clustering algorithm for classifying errors and monitoring defect evolution.
    \item We construct a performance metric for log evolution based on the experience of real SREs.
    \item We demonstrate that our algorithm improves upon past clustering techniques on a private dataset and provide baseline metrics on public log datasets.
\end{enumerate}

\begin{figure*}
    \captionsetup[subfigure]{labelformat=empty}
     \centering
     \begin{subfigure}[t]{0.28\textwidth}
         \centering
         \includegraphics[width=\textwidth, height=6.5cm]{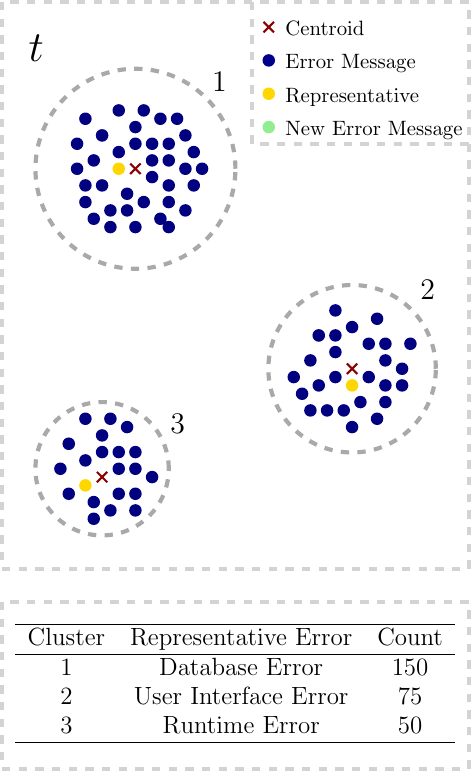}
         \hspace*{\fill}
         \caption{Time $t$}
         \label{fig:problem statement t0}
     \end{subfigure}
     \hskip1em
     \begin{subfigure}[t]{0.28\textwidth}
         \centering
         \includegraphics[width=\textwidth, height=6.5cm]{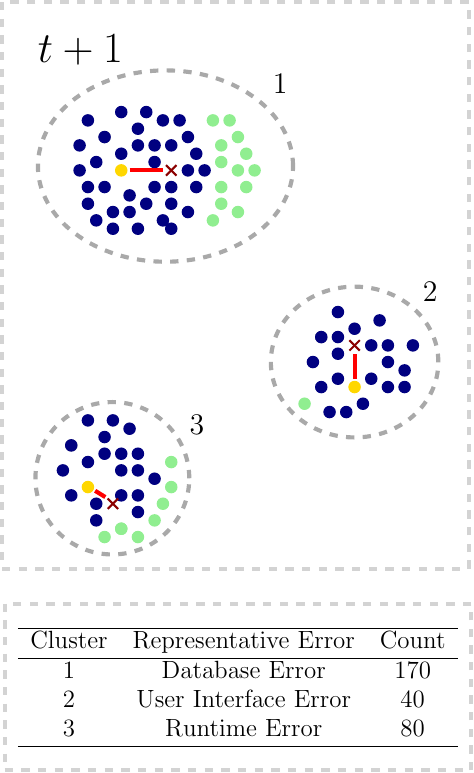}
         \hspace*{\fill}
         \caption{Time $t+1$}
         \label{fig:problem statement t1}
     \end{subfigure}
     \hskip1em
     \begin{subfigure}[t]{0.28\textwidth}
         \centering
         \includegraphics[width=\textwidth, height=6.5cm]{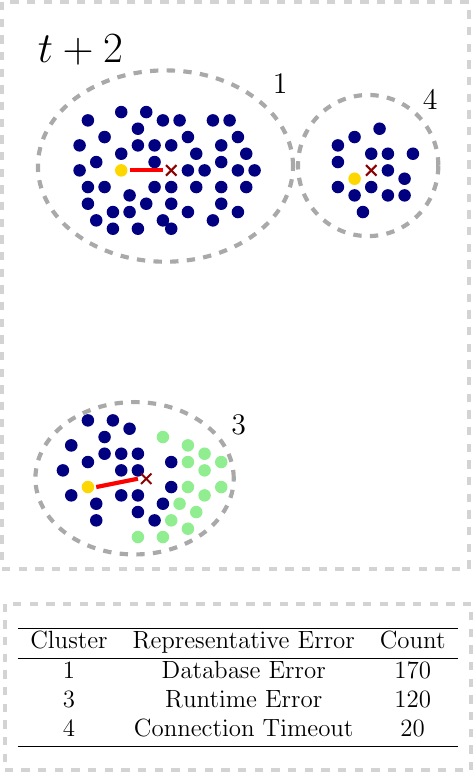}
         \hspace*{\fill}
         \caption{Time $t+2$}
         \label{fig:problem statement t2}
     \end{subfigure}
        \caption{Example of Error Log Evolution. The top row provides a visual aid of the clusters while the bottom row details the specific defects found at each point in time.}
        \label{fig:problem statement}
\end{figure*}

\section{Related Work}
\label{Related work}

Log mining in the context of software defects is a well-researched topic~\cite{cyber_survey} which mainly focuses on failure diagnosis~\cite{failDiag}, anomaly detection~\cite{anom} and log comprehension~\cite{logComprehension}. Older research focuses on tracking log patterns over time through heuristics like frequency, sentence length or n-grams to group logs together~\cite{rule_based1, LenMa, logram}.
More recently, the FLAP system was introduced as an end-to-end framework for log analysis \cite{FLAP} based on syntactic log properties and an iterative clustering process \cite{IPloM}.
Other work exists that utilises syntactic features to group logs \cite{LKE, LogMine,LogCluster,LogSig} and track formed clusters \cite{DRAIN, Spell}.
The techniques presented in these works all fail to extract a representative log for clustering which is a desirable quality in terms of the explainability of the log error types.
Similar work has shown that it is possible to achieve syntactic representatives \cite{automatic_detection, starlink, towards_detecting_patterns}. 
Furthermore, past work has also utilised genetic algorithms to extract log templates \cite{Genetic}.

Semantic representations of logs have been explored in anomaly detection and defect prediction \cite{deep_log, swiss_log, log_robust, log_anomaly}.
Research also exists that examines log comprehension and clustering using semantic representations (e.g., Word2Vec) \cite{clust_error_mess}.
This work differs from the system proposed in this paper in two fundamental ways.
Firstly,  it treats logs using a bag-of-words (BoW) approach and so does not consider word order nor frequency.
Therefore, it discards important information that exists in more complex representations such as the sentence embeddings of SBERT \cite{SBERT}.
Secondly, it applies the DBSCAN clustering algorithm which is a static algorithm and so it is impossible to track log clusters over time.
There are streaming clustering techniques such as DenStream \cite{denStream} which work with vector representations, however, they do not provide representative extraction and can have an erratic convergence.

\section{Tracking Error Log Evolution}
This paper concerns the monitoring of error logs which are descriptive of various defects encountered in a system's execution.
We consider a \textbf{log} to be an individual message sent by the system during its execution (logs sent from the same execution are related by an execution ID).
In order to motivate and devise an effective solution to our task, we introduce three key criteria for successful monitoring of log evolution that were reached in collaboration with over 60 SREs.
We further motivate our criteria by a visual example in \cref{fig:problem statement}.

\paragraph{Identifying Errors over Time} We wish to correctly discover patterns that represent a common defect or error expressed by logs. We expect the types of errors to change over time and be correctly captured by the algorithm. As can be seen in \cref{fig:problem statement}, over time, we no longer witness ``User Interface Error`` defects while we begin to witness ``Connection Timeout`` defects.

\paragraph{Capturing Meaningful Representatives} It is important that we are able to describe each cluster by its common defect. As such, a successful algorithm should provide a meaningful representative to each cluster that an end-user is able to understand. A meaningful representative is visually depicted to be near the centre of each cluster in \cref{fig:problem statement}. Additionally, each cluster has an understandable and succinct representative error that aligns with common code defects as seen in the tables of \cref{fig:problem statement}.

\paragraph{Minimising Disruption} In order to effectively monitor the lifecycle of an error or defect, clusters must remain active until their associated errors are fully resolved.
As such, we require clusters to evolve smoothly over time while maintaining high quality clusters.
There is minimal disruption illustrated in \cref{fig:problem statement} as cluster 1 expands and eventually breaks off into cluster 4 (perhaps along with cluster 2).
The smooth evolution over time is needed to not introduce or remove a defect cluster too quickly or erratically.

\subsection{A Novel Online Clustering Algorithm}
One of the contributions of this work is a novel online clustering algorithm for error logs that is motivated by the above criteria.
We draw some inspiration from DenStream \cite{denStream} who create dynamically changing clusters.
In particular, we adopt the same distance metric between logs and cluster centroids.
The intuition of our online clustering algorithm is straight-forward:
When new data points are received, we treat points either one by one or in batches and merge them into the cluster with the smallest cosine distance to the centroid of that cluster.
We formulate our algorithm in \cref{alg:sam} using three hyperparameters:
\begin{itemize}[noitemsep, leftmargin=0.2cm]

\item[] $\theta$: The acceptance threshold of a log to a cluster.
If the distance between a log and all cluster centroids is less than $\theta$, a new cluster is created.

\item[]$\alpha$: The rate of cluster centroid evolution.
We update cluster centroids as a rolling average based on new logs that are merged into a cluster. Consequently, cluster centroids naturally change over time. To prevent this from occurring too quickly (and thus minimise disruption), we use $\alpha$ to slow change.

\item[]$\gamma$: The minimum cluster size before a centroid of a cluster may change. We use this hyperparameter to ensure outliers do not heavily impact smaller clusters that are more liable to centroid changes.

\end{itemize}

\begin{algorithm}[t]
\KwData{List of batches}
\KwInp{$\theta$; $\alpha$; $\gamma$}
clusters = []\;
\For{batch \KwIn data}{
    \For{p \KwIn batch}{
        min\_dist = inf; c = null\;
        \For{clust \KwIn clusters}{
            dist = 1 - cosine\_sim(clust.cen, p)\;
            \If{dist $<$ min\_dist}{
                min\_dist = dist\;
                c = clust\;
                }
        }
        \eIf{min\_dist $\le$ $\theta$}{
             \eIf{c.len $\ge$ $\gamma$}{
              c.cen = (1-$\alpha$) $\cdot$ c.cen + $\alpha p$\;
             }{
        c.cen = $\frac{c.len}{c.len+1} \cdot c.cen$ + $\frac{1}{c.len + 1}p$\;
        }
        c.len += 1
        }{
        clust = new Clust(cen=p, len=1)\;
        clusters.add(clust)\;
        }
    }
}
\caption{Our online clustering algorithm, with updates coming as individual points. Batched updates are more efficient.}
\label{alg:sam}
\end{algorithm}

For an end user to clearly interpret the clustering results, we extract an error message from each cluster.
For each cluster, we use cosine similarity to select the sample which is closest to the centroid and use its log as the cluster's representative defect. 
Alternatively, the Levenshtein score can be used to get the average message \cite{Lev_dist}.
While this method may give a better message as it is more specific to the log text, it is more computationally expensive and so we stick to the cosine similarity approach.\footnote{Let each cluster have $m$ logs, the average log length is $n$, and the semantic representation dimensionality is $d$ where $d << n$. Computing the Levenshtein score has a time complexity of $O(m^2n^2)$  as opposed to  $O(md)$ for the cosine similarity approach.}

\subsection{Performance Metrics}
We next introduce a novel metric that formalises the three key criteria identified at the start of this section.
Our metric is devised of three scores: Silhouette score ($S$), Representative Similarity ($R$) and Number of Clusters Similarity ($C$).
Each of these scores is computed over batches and is in the range $[0,1]$.

The Silhouette score is a standard metric used for assessing unlabelled clustering.
It provides a measure of the similarity of each log to its own cluster versus other clusters.
Traditionally, the score is given in the range $[-1, 1]$, but we scale it so that it is comparable with our other scores.
\begin{equation}
     S = \frac{1}{B} \sum_{b=1}^B  \frac{Silhouette_b + 1}{2}
\end{equation}
where $Silhouette_b$ is the Silhouette score for bath $b$.

The representative similarity score compares the representatives of each cluster $c$ for batches $b$ and $b+1$ coming in at times $t$ and $t+1$ respectively. 
We use this score to evaluate the evolution of the representative logs for each cluster.
We use the cosine similarity as our distance metric. 

\begin{equation}
        R = \frac{1}{B} \sum_{b=1}^B cosine\_sim(repres_{b}, repres_{b+1})
\end{equation}
where $repres_{b}$ is the semantic representations for batch $b$.

Lastly, the number of clusters similarity score measures how smoothly the number of clusters evolves over time.
\begin{equation}
     C = 1 - \frac{1}{B} \sum_{b=1}^B  \frac{\left\lvert nr\_clust_{b} - nr\_clust_{b-1} \right\rvert}{max(nr\_clust_{b}, nr\_clust_{b-1})}
\end{equation}
where $nr\_clust_{b}$ is the number of clusters in batch $b$.

Our log cluster evolution metric ($LCE$) is then comprised of a linear combination of the three scores above, where all scores are scaled to be in the range $[0,1]$ and  $w_S + w_R + w_C = 1$
\begin{equation}
    LCE=  w_S S + w_R R  + w_C C
\end{equation}

\section{Experiments}
In this section, we measure the performance of our online clustering algorithm using different semantic representations.

\subsection{Data}
We primarily conduct experiments on a private industrial dataset from the SREs we worked with. 
To do this, we collected log data from a database containing all monitored applications.
We then eliminated variability by removing timestamps and URLs.
The monitored applications generate around $100,\!000$ logs per day, of which, about $1.5\%$ are related to errors and defects.
We extracted two months' worth of data, resulting in a total of $57,\!000$ error logs.
Each log is comprised of structured fields (e.g., system id, date, log level) as well as unstructured, free-form text.

Since our dataset is private and cannot be shared, we also evaluate our method on similar public datasets. We use Loghub~\cite{LogDataset}, a collection of logs from different systems which have a date element, a log level and free-form text representing the meaning of the log.
We note that this dataset is significantly simpler than the private industry dataset due to generally shorter and less varied log texts.
We use a sample of $2,\!000$ logs for each of the Loghub systems: HDFS\_2, Linux, Zookeeper and OpenStack.
Note that each dataset is much smaller than the private dataset and examines a single framework rather than many applications.
Throughout both the public and private datasets, we consider a log to be stale after one month since its creation.

\paragraph{Pre-processing and Semantic Representations}
Logs are a challenging data type for text understanding models because they tend to contain incomplete words, words with special characters (e.g., underscore), and system-specific terms.
Consequently, we cleaned the data using standard natural languages processing techniques, such as tokenization\cite{tokenisation}, lemmatization\cite{lemantisation}, and removal of English stop words given by the SpaCy library\footnote{https://spacy.io/}. 
Given the cleaned logs, we constructed semantic representations of the logs using average Word2Vec\footnote{https://spacy.io/usage/spacy-101/\#vectors-similarity} vectors as well as SBERT sentence embeddings \cite{SBERT}.

\paragraph{Temporal Splits}
For the private dataset, we first split the two months' data into a one-month batch and the rest into five-day batches. The second split considers only one-day batches. The reason for these two separate temporal settings is to show results on stationary versus dynamic data.
In the first split, due to the high amount of data in the snapshot, the underlying distribution does not differ from the next timestamps. However, for one-day splits, one batch is not a meaningful sample of the entire distribution of logs.  We use these datasets to compare our model with other clustering algorithms.
We use one day batches for the public datasets.

\subsection{Models and Algorithms}
To benchmark our algorithm against the literature we picked two well-known models which work with vector representations. 
Gaussian Mixture Model (GMM) is an offline clustering algorithm which was selected because it allows clusters of any shape to be formed.
We adapt the algorithm to be online by re-initialising each batch with the mean and covariance of the model trained on the previous batch.
We choose DenStream \cite{denStream} as the second algorithm because it trains in an online fashion and also allows clusters of varying shapes. The optimal values for the hyper-parameters of all models were selected based on a grid search conducted over past data.\footnote{For our algorithm, the hyperparameters were $\theta=0.05$, $\alpha=0.1$, and $\gamma=100$.}
For each algorithm we only use Word2Vec (embedding dimension of $300$) to compare the algorithm's performance.
We further use SBERT (embedding dimension of $756$) for our algorithm to show the improvement in using a more complex and rich semantic representation.
Non eof the models were fine-tuned on the data.

\begin{table}[t!]
\center
  \caption{Clustering evolution performance on private dataset.
  }%
\begin{tabular}{@{}lccccc@{}}
\midrule
\multicolumn{1}{l}{\textbf{Model}} & \textbf{\#\ Clusters} & $\boldsymbol{R}$ & $\boldsymbol{S}$ & $\boldsymbol{C}$ & $\boldsymbol{LCE}$ \\ \midrule

& \multicolumn{5}{c}{\textbf{1 Month snapshot + 5 day batches}} \\
\cline{2-6} \\
GMM$_{\mathrm{Word2Vec}}$                     &  11 & 0.964 &  0.768 &  \textbf{1} & 0.9 \\
DenStream$_{\mathrm{Word2Vec}}$              &  48 & 0.92 & 0.771 &   0.803 & 0.83 \\ 
Ours$_{\mathrm{Word2Vec}}$               &  198 & 0.999 & 0.865 &  0.96  & 0.94 \\
Ours$_{\mathrm{SBERT}}$       &  27 & \textbf{0.999} & \textbf{0.97} & \textbf{ 0.96}  & \textbf{0.97} \\
\midrule

& \multicolumn{5}{c}{\textbf{1 day batches}} \\
\cline{2-6} \\

GMM$_{\mathrm{Word2Vec}}$       &  11 & 0.914 &  0.727 &  1 & 0.88 \\

DenStream$_{\mathrm{Word2Vec}}$         &  48 &  0.92 & 0.771 &  0.862 & 0.85 \\ 
Ours$_{\mathrm{Word2Vec}}$              &  205 & \textbf{0.999} & 0.879 & \textbf{0.999} & 0.95 \\
Ours$_{\mathrm{SBERT}}$            &  28 & \textbf{0.999} & \textbf{0.972} & \textbf{0.998} & \textbf{0.98} \\
\midrule
\end{tabular}
\label{tab:results table}
\end{table}

\begin{table}[t]
    \caption{Clustering evolution performance on private and public datasets using Ours$_{\mathrm{SBERT}}$ and one day batches.}%
    \begin{tabular}{@{}lccccc@{}}
    \midrule
    \textbf{Dataset} & \textbf{\#\ Clusters} & $\boldsymbol{R}$ & $\boldsymbol{S}$ & $\boldsymbol{C}$ & $\boldsymbol{LCE}$  \\
    \hline
        Private Dataset &  28 & {0.999} & 0.972 & {0.998} & 0.980 \\
        HDFS\_2 & 18 & 0.990 & { 0.990} & 0.990 & {0.990}\\
        Linux & 31 & 0.990 & 0.860 & 0.980 & 0.940\\
        Zookeper & 19 & 0.990 & {0.990} & 0.990 & {0.990}\\
        OpenStack & 14  & 0.990 & 0.790 & 0.990 & 0.920  \\
        \bottomrule
    \end{tabular}
    \label{tab:pub_data}
\end{table}

\subsection{Results}

For all results, $LCE$ was computed with equal contribution weights.
\cref{tab:results table} shows the performance results across the different algorithms and semantic representations on the private dataset.
We observe that for both temporal settings, our approach matches or outperforms both the GMM and DenStream algorithms when using Word2Vec semantic representations across each individual score as well as $LCE$.
We note that GMM, the static clustering algorithm, performs better when it has the initial one month snapshot whereas the dynamic approaches perform consistently across both settings.
This is expected as GMM requires the initial mean and variance to attain better performance.

The biggest improvement between our approach DenStream is observed in the cluster number similarity score.
DenStream exhibits an erratic evolution which we are able to mitigate in our approach using our evolution rate ($\alpha$) and minimum cluster size hyperparameters ($\gamma$).
The strong and consistent performance of our approach across both temporal settings suggests that our algorithm adapts easily to both stationary and non-stationary streams.
The robustness of our algorithm is further shown in \cref{tab:pub_data} where we achieve good results across public log datasets.\footnote{We further conducted a brief qualitative analysis with the SREs we worked with. They provided positive feedback with regards to the cluster evolution over time and pointed out certain clusters matched specific errors they were aware of.}

\section{Threats to Validity}
We note a few threats to validity in this work.
Firstly, we must be cautious of overfitting to our data which would retract from the generalisibility of the model.
As clustering is an unsupervised problem (i.e., no labelled data), we cannot use metrics such as ARI to evaluate our predictions.
While the Silhouette score and manual qualitative analysis can be done to demonstrated good clusters,
having access to a small labelled dataset would allow for more robust evaluation.
This is currently a gap in log-based software defect datasets that we hope is addressed in future work.
Indeed, one could use a model as presented in this work to generate data.

\section{Conclusion}
In this paper, we introduced a novel online semantic-based clustering algorithm for error logs.
Our algorithm is able to cluster log message streams, track the reduction and emergence of defects that arise in code, and provide a representative defect message for each cluster. 
The algorithm has configurable cluster granularity and cluster evolution rate through hyperparameters. 
Furthermore, we introduced a novel metric named $LCE$ that can evaluate defect tracking performance based on industry specified criteria.
We demonstrated that our algorithm outperforms existing log clustering approaches using a private dataset and showed strong performance across public datasets.
We hope that this work encourages more work in defect detection to examine the temporal properties of logs and incorporate it into modern systems.

\section*{Disclaimer}
This paper was prepared for informational purposes by the Artificial Intelligence Research group of JPMorgan Chase \& Co and its affiliates (“JP Morgan”), and is not a product of the Research Department of JP Morgan. JP Morgan makes no representation and warranty whatsoever and disclaims all liability, for the completeness, accuracy or reliability of the information contained herein. This document is not intended as investment research or investment advice, or a recommendation, offer or solicitation for the purchase or sale of any security, financial instrument, financial product or service, or to be used in any way for evaluating the merits of participating in any transaction, and shall not constitute a solicitation under any jurisdiction or to any person, if such solicitation under such jurisdiction or to such person would be unlawful.

\bibliographystyle{ACM-Reference-Format-New}
\bibliography{references}

\end{document}